# A novel technique to characterize the spatial intra-pixel sensitivity variations in a CMOS image sensor

Swaraj Mahato[a], J. De Ridder[a], Guy Meynants[b], Gert Raskin[a], H. Van Winckel[a]

[a]Institute of Astronomy, KU Leuven, Celestijnenlaan 200D, 3001 Leuven, Belgium;  [b] AMS, Coveliersstraat 15, 2600 Antwerpen, Belgium

Email: swarajbandhu.mahato@kuleuven.be

*Abstract*— To understand the scientific imaging capability, one must characterize the intra-pixel sensitivity variation (IPSV) of the CMOS image sensor. Extracting an IPSV map contributes to an improved detector calibration that allows to eliminate some of the uncertainty in the spatial response of the system. This paper reports the measurement of the sub-pixel sensitivity variation and the extraction of the 2D IPSV map of a front-side illuminated CMOS image sensor with a pixel pitch of 6 μm. Our optical measurement setup focuses a collimated beam onto the imaging surface with a microscope objective. The spot was scanned in a raster over a single pixel and its immediate neighbors in order to probe its response at selected (sub-pixel) positions. In this work we introduced a novel technique to extract the IPSV map by fitting (forward modeling) the measured data to a mathematical model of the image, generated in a single pixel that allows for a spatially varying sensitivity.

## I. Introduction

Spatial variations of the sensitivity of a solid state imager influences the total flux detected by the imager with respect to the position of the point source within the pixel. The sensitivity variation inside a single pixel is called intra-pixel sensitivity variation (IPSV) or intra-pixel response non-uniformity (IRNU) [1]. Scientific applications like space missions are designed for high-precision photometry where the IPSV is important. It causes an extra source of noise in the integrated flux signal because of the slight movement of the point spread function (PSF) within a pixel due to spacecraft pointing jitter. Intra-pixel sensitivity profiles of the front as well as back illuminated CCD's were studied extensively [2]-[5] but a few works have been done on CMOS image sensor [6]. To understand the scientific imaging capability, one must therefore characterize the intra-pixel sensitivity variation of the CMOS image sensor. This paper reports the measurement of the sub-pixel sensitivity variation and the extraction of the 2D IPSV map of a front-side illuminated CMOS image sensor with a pixel pitch of 6 μm. Our optical measurement setup focuses a collimated beam onto the imaging surface with a microscope objective. The spot was scanned in a raster over a single pixel and its immediate neighbors in order to probe its response at selected (sub-pixel) positions. The signal measured by a single pixel as a function of the beam spot position, can be modeled by the convolution of the iPSF and the IPSV. Most of the previous work in the literature derives the IPSV by performing a de-convolution (backward modeling) in which the measurement uncertainty is not taken into account. In this work we introduced a novel technique to extract the IPSV map by fitting (forward modeling) the measured data to a mathematical model of the image generated in a single pixel that allows for a spatially varying sensitivity. We minimize the measure of the goodness of fit (chi-square) between the observed data and the model using a modified version of the non-linear Levenberg-Marquardt algorithm, and present the calculated IPSV map. Unlike the traditional de-convolution method our forward modeling method includes the measurement uncertainty and the crosstalk.

Section II describes our measurement setup and the overview of the measurement technique. Intra-pixel sensitivity computation using our novel forward modeling method with pixel response fitting is described in section III. The results from the IPSV extraction technique is presented in section IV. Finally, section V concludes the paper.

## II. Measurement Details

### A. Measurement Setup

In order to study the intra-pixel sensitivity of the CMOS image sensor in lab, we used an optical setup to project a very small light spot onto the detector surface. The setup focuses a collimated beam onto the image sensor by a microscope objective. All measurements were performed with a pinhole of 20 μm as the object. The light source is a 250 W incandescent lamp powered by a Statron Typ 3202 DC power supply. The lamp is mounted inside a housing which contains a lens at its exit port. The lamp is positioned such that the image of the glowing filament is in focus at the object. The spot was scanned in a stepped raster fashion over a single pixel and its immediate neighbors in order to probe its response at selected (sub-pixel) positions. The system includes a reverse telescope and a microscope objective. A motion controller with 3 controllable axes (X, Y and Z) can be programmed to scan the image sensor over a certain range of pixels. The optical measurement setup and its schematic are shown in fig 1. The system we used is diffraction limited. This means that the size of the beam spot is limited only by the laws of physics and not by aberrations of the lenses. Because of the circular shape of the microscope objective in our measurement, we consider the beam profile as an Airy diffraction pattern, which defines the instrumental point spread function (iPSF) of the optical setup. The microscopic objective, used for this measurement is Carl Zeiss GF-Planachromat 12.5x. The numerical aperture of this microscope objective is NA=0.25. The Abbe limit is given by:

$$r = \frac{1.22\,\lambda}{2n \cdot sin\theta} = \frac{1.22\,\lambda}{2 \cdot NA} \qquad (1)$$

Where r is the radius of the spot (distance from centre to first minimum of the Airy disc), λ is the wavelength, n is the refractive index of the lenses, θ=tan(d/(2f)) is the converging angle defined by lens diameter d and focal length f, and NA is the numerical aperture. The calculated radius of the spot of our setup is 1.34 μm (used green filter, λ = 550 nm).

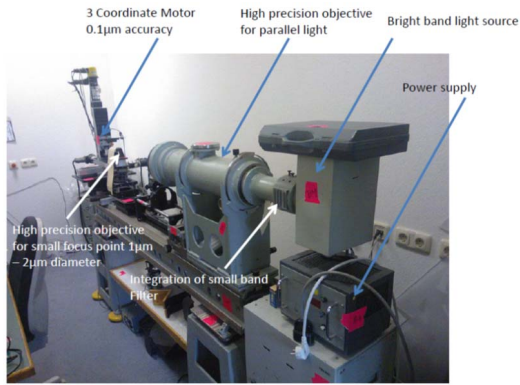

(a)

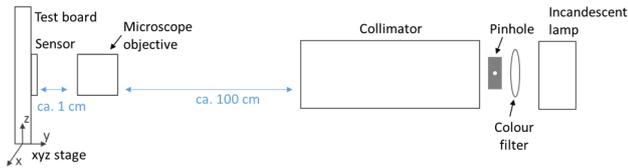

(b)

Fig. 1.  (a) Optical measurement setup. (b) Schematic of the setup.

### B. Scanning Geometry of The CMOS Front-side Illuminated Image Sensor

Our image sensor is a CMOS front-side illuminated image sensor with a pixel pitch of 6 μm. The spot is focused to the certain pixel, which will be the pixel of interest (POI) for each measurement. According to the optical setup, before start the scan we need to position the optical spot to the centre of the POI and then moved to the starting position (upper left corner) of the scan area. As we don't have any automatic system, the centre of the POI is obtained by finding the position where neighboring pixels get the equal amount of pixel count (light). After that the Y-axis of the motion controller (collinear one with the optical axis) was adjusted to set the detector in the focus plane of the microscope objective. This process attempts to find the position where all neighboring pixels get the least amount of light (pixel count).  Both centre positioning and focusing is done with multiple iterations when images were grabbed continuously. Fig. 2 depicted the focused POI with the optical spot is positioned at the centre.

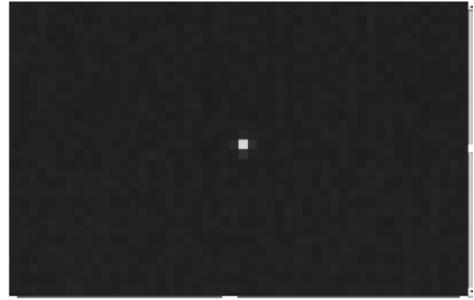

Fig. 2.  Focused optical spot at the centre of the single pixel.

The sensor was scanned over a 5 × 5 pixel area. The scans were performed by drifting the beam across the detector and capturing multiple sub-frame images for each spot position. The use of sub-frame readout increased the speed of the scanning process and allowed to get the average of the many scans in order to increase the signal-to-noise ratio. For the sub-frame image we defined 9x9 pixel area around the POI for which data will be captured. The scanning geometry is depicted in fig 3. The pixel size of the detector is 6μm x 6μm and the scanning step-size is selected as 0.6μm. So, a single scan line is comprised of 10 steps per pixel for a total of (50+1) scan points on total 5 pixels. At each scan point we carefully selected the exposure time to keep a high signal-to-noise ratio while avoiding saturation (<16000 ADU) and nonlinearity.

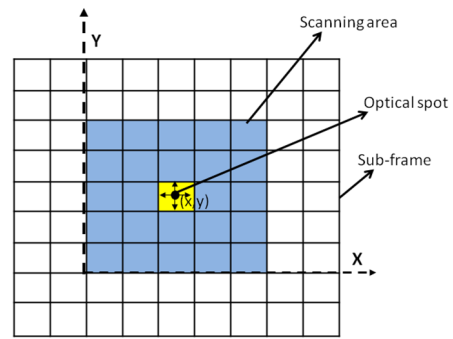

Fig. 3. Scanning geometry for the experiment. The optical spot (dark circle/dot) is scanned over the array in a raster fashion within the scanning area (blue).

For each scan point, we captured 12 sub-frames and calculated the averaged frame to reduce noise. All averaged frame of the scans in a raster sequence were stored in data files on a computer. The scanning area contains 2601 (=51x51) measurements in total. Later analysis was performed with PYTHON script. All files were first read out and the middle pixel's ADU value is stored in a new matrix (51 x 51). Storing the data was done in the same order as the scanning was executed. The operating temperature was around 24˚C in the dark room. The obtained signal of the POI as a function of the spot position provides the pixel response image, shown in Fig. 4.

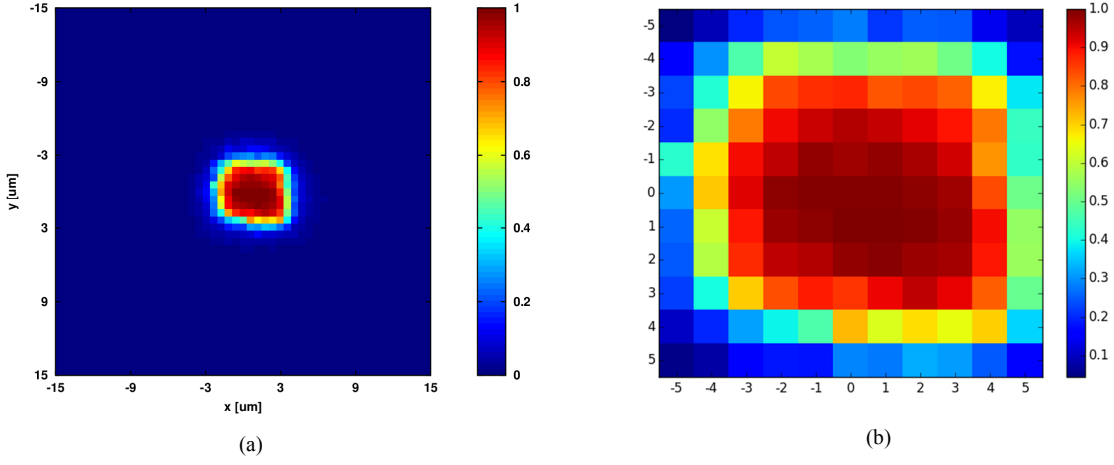

Fig. 4. (a) Pixel (POI) response as a function of the spot positions over the full scan area. (b) Pixel (POI) response as a function of the sub-pixel spot positions on the POI.

## III. INTRA-PIXEL SENSITIVITY VARIATION COMPUTATION

Deriving the pixel response in the image plane is a task similar to find a mathematical model of image generated in a single pixel. This can be derived by using the beam spot profile over the POI. The beam spot profile is described in the form of airy diffraction pattern, defined as:

$$I(\theta) = I_0 \left(\frac{2J_1(ka\sin\theta)}{ka\sin\theta}\right)^2 \quad (2)$$

$$I(x) = I_0 \left(\frac{2J_1(\rho)}{\rho}\right)^2, \text{ with } \rho = ka\sin\theta \quad (3)$$

Here $I_0$ is the maximum intensity of the pattern at the Airy disc centre, $J_1$ is the Bessel function of the first kind of order one, $k=\frac{2\pi}{\lambda}$ is the wave-number, $a$ is the radius of the aperture, and $\theta$ is the angle of the observation. Fig.5 is the 2D representation of the calculated spot profile (Airy disk) for our microscope objective with NA=0.25.

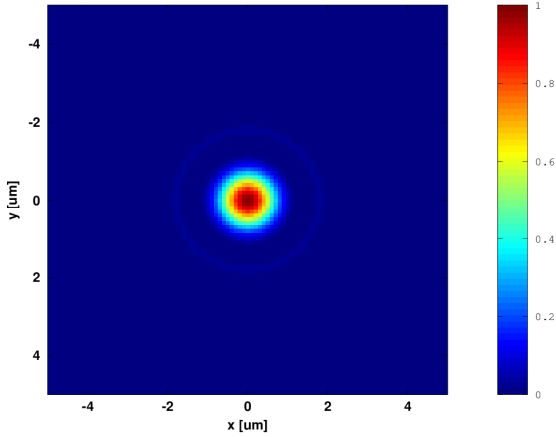

Fig. 5. Spot profile (2D Airy function).

In Cartesian coordinate, $\rho \equiv \frac{w_0}{w}\sqrt{(x^2+y^2)}$. Where $w_0$ is the location of its first zero ($J_1(\rho) = 0$) and $w$ is the radius of the input image which is calculated as 3.14 μm at 550 nm for the used microscope objectives (NA = 0.25). When normalized, $I_0 = 1$ then,

$$I(x,y;w;\lambda) = \left(\frac{2J_1\left(\frac{w_0}{w}\sqrt{(x^2+y^2)}\right)}{\frac{w_0}{w}\sqrt{(x^2+y^2)}}\right)^2 \quad (4)$$

We defined a sub-pixel bin of 5x5 for the single POI. The sensitivity of the each sub-pixel is defined by $S_{ij}$ where (i,j) is the sub-pixel spatial coordinate. Now, let define $model(x_k,y_k;S_{1,1},S_{1,2},S_{1,3},....S_{5,5})$ be a model of the intensity values that has two spatial coordinates $(x_k,y_k)$ of the spot position and 25 sub-pixel sensitivity parameters. For notational convenience, let the vector $r_k$ represents the coordinates $(x_k,y_k)$ and the vector $S_k$ represents all the parameters [i.e., $S_k \equiv (S_{1,1},S_{1,2},S_{1,3},....S_{5,5})$]. Thus the model of intensities will normally be written as $model(r_k;S_k)$. Then the model of the pixel response for each observation is defined as:

$$model(r_k;S_k) = model(x_k,y_k;S_{ij})$$

$$= \sum_{i,j}^{5x5} S_{ij} \int_{ymin,j}^{ymax,j} \int_{xmin,i}^{xmax,i} I(x-x_k,y-y_k;w;\lambda)dx\,dy \quad (5)$$

$(x_k,y_k)$ is the spot position in each observation which represents the sub-pixel offset of the input with respect to the pixel grid. ($ymin,j$ & $ymax,j$) and ($xmin,i$ & $xmax,i$) define the boundaries of the sub-pixel $(i,j)$. The IPSV is the sensitivity map ($S_k$) for the sub-pixel regions which minimizes the measure of the goodness of fit (chi-square) between the observed data from the measurements and the model. It is defined as:

$$\chi^2 = \sum_{k=1}^{N}\left[\frac{1}{\sigma_k}(Obs_k - model(r_k;S_k))\right]^2 \quad (6)$$

Here, $N$ is the number of pixel intensity values or the total spot scan points over the POI and $\sigma_k$ is the uncertainty of the measurement $Obs_k$. The observed data, $Obs_k$ is the image, captured for $k^{th}$ spot position. Solution of the "least squares" fitting will give the sensitivity map. The PYTHON routine is developed based on the modified version of the non-linear Levenberg-Marquardt algorithm, used for the fitting [7]. We select $\sigma_k$, the uncertainty of the measurement $Obs_k$ to weight the fit as $1/\sigma_k = 1/\sqrt{Obs_k}$. The routine requires an initial set of sensitivity parameters which is modified until a good fit is achieved. We select all initial sub-pixel sensitivity as 100%.

## IV. RESULTS

The theory of least squares states that the optimum value of the parameter vector $S_k$ is obtained when $\chi^2(S_k)$ is minimized with respect to each parameter simultaneously. Data fitting of the model with observation is shown in fig. 6 where in the x-axis of this 1d plot is the consecutive chain of scan points (total 11x11=121 scan points) of the raster sequence. This figure shows that the calculated POI value (model's output) is fitted well with the observed POI value for each sub-pixel spot position.

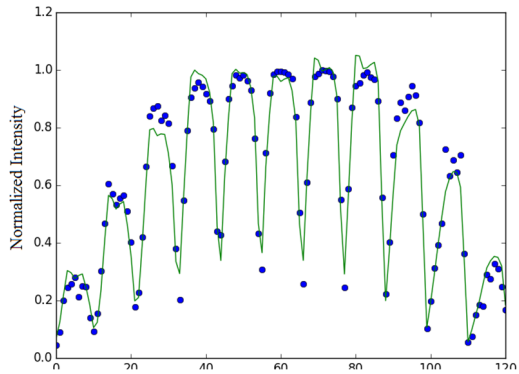

Fig. 6. Data fitting of the model with observation.

In our work if the function $\chi^2(S_k)$ is anticipated as a surface in the 25 dimensional parameter space (5x5) and if $S_{k,true}$ is defined as the optimal parameter vector, then the absolute minimum of that surface is $\chi^2(S_{k,true})$. The method to minimize the squares is commonly called the method of "least squares". Solution of the "least squares" fitting will give the optimal parameter vector which is the calculated intra-pixel sensitivity variation map or pixel response function, presented in fig. 7. This intra-pixel sensitivity variation map shows how the sensitivity of the POI varies across the sub-pixel grid.

For scientific application a high performance imaging system can be calibrated by mapping the response of the image sensor for different positions of the optical point spot within a single pixel. In space astronomy, detector calibration using IPSV, contributes to the high precision altitude determination as well as to the target tracking across the focal plane.

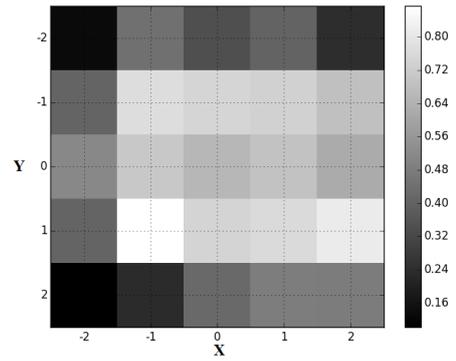

Fig. 7. 2D Map of the calculated intra-pixel sensitivity variation or pixel response function of the POI.

## V. CONCLUSION

In this paper, a novel technique to extract the intra-pixel sensitivity variation map for the CMOS image sensor is introduced. This forward modeling technique fits the measured data to a mathematical model of the image, generated in a single pixel that allows for a spatially varying sensitivity. This technique also includes the measurement uncertainty. The details of the measurement setup and the procedure have been described. Results show the good fitting of the data with model output and the extracted IPSV map of the CMOS image sensor. The extracted IPSV map can then be used as a calibration filter which would contain the information related to the overall shape of a point source as viewed through an optical system by the image sensor.

### *Acknowledgment*

This work is funded by IWT-Baekeland.